\documentclass[%
reprint,
superscriptaddress,
amsmath,amssymb,
aps,
prl,
floatfix,
]{revtex4-1}
\usepackage{graphicx}
\usepackage{bm}
\usepackage{amsmath}
\usepackage{dcolumn}
\newcolumntype{d}[1]{D{.}{.}{#1}}

\newcommand{\qw}{\phantom{$-$}}

\newcommand{\bol}[1]{\mbox{\boldmath $#1$}}
\newcommand{\be}{\begin{equation}}
\newcommand{\ee}{\end{equation}}

\def\v{\bol}

\newcommand{\RR}[0]{\mathcal{R}}

\begin{document}

\title{Frequency-dependent polarizability
of helium including relativistic effects with nuclear recoil terms
}

\author{Konrad Piszczatowski}
\affiliation{Department of Physics and Astronomy, University of Delaware, 
             Newark, DE 19716}

\author{Mariusz Puchalski}
 \affiliation{Faculty of Chemistry, Adam Mickiewicz University,
             Umultowska 89B, 61-614 Pozna\'n, Poland}

\author{Jacek Komasa}
\affiliation{Faculty of Chemistry, Adam Mickiewicz University,
            	Umultowska 89B, 61-614 Pozna\'n, Poland}

\author{Bogumi{\l} Jeziorski}
\affiliation{Department of Chemistry, University of Warsaw,
             Pasteura 1, 02-093 Warsaw, Poland}

\author{Krzysztof Szalewicz}
\affiliation{Department of Physics and Astronomy, University of Delaware, 
             Newark, DE 19716}
 
\date{\today}

\begin{abstract}
Future metrology standards will be partly based on physical
quantities computed from first principles rather than measured.  In particular,
a new pressure standard can be established if the dynamic polarizability
of helium can be determined from theory with an uncertainty smaller than
0.2 ppm.  We present calculations of the frequency-dependent part of
this quantity including relativistic effects with full account of
leading nuclear recoil terms and using highly optimized explicitly
correlated basis sets.  A particular emphasis is put on uncertainty
estimates.  At the He-Ne laser wavelength of 632.9908 nm, the computed polarizability
value of $1.391\,811\,41$ a.u.~has uncertainty of 0.1 ppm that is two
orders of magnitude smaller than those of the most accurate polarizability
measurements.  We also obtained an accurate expansion of the helium
refractive index in powers of density.
\end{abstract}

\maketitle



Some physical quantities, for example, properties of the helium atom and
interaction energies of helium atoms, can now be computed from first
principles with precision rivaling and sometimes  exceeding the best
experimental determinations \cite{Lach:04,Przybytek:10,Cencek:12}.
Therefore, quantities of this type can be used in establishing metrology
standards.  One example is a possible standard of temperature based on
acoustic gas thermometry \cite{Moldover:14}. Another example is a
pressure standard based on optical interferometry \cite{Hendricks:14a}.
The current pressure standard dating back more than 300 years is realized by
mercury manometers and can not be further improved.  Also, the reference
manometers are far from portable: 3 m high and containing 250 kg of
mercury, a substance banned due to its toxicity.  Since pressure is one
of the most widely measured properties, in applications ranging from
manufacturing of semiconductor chips to air-traffic control, a new
pressure standard would significantly impact both technology and
everyday life.  The proposed standard \cite{Hendricks:14a}
obtains pressure from the formula
\cite{Schmidt:07}
\begin{equation}\label{pressure} 
p = \frac{n^2 -1}{n^2 +2}\,\,
           \frac{3kT}{ 4\pi(\alpha  + \chi ) } + \cdots,
\end{equation}
where $n$ denotes the index of refraction of helium gas,
$k$ the Boltzmann constant, $T$ the temperature,  $\alpha$
the dipole polarizability, and $\chi$ the diamagnetic susceptibility of
helium.  To account for nonideality of helium
gas,  one has to include some small terms on the right-hand side
depending on dielectric and density virial coefficients
\cite{Schmidt:07}.  The essential part of the new standard is the
determination of $n$ with an uncertainty of 0.2 ppm via interferometric
measurements of a variable-length cavity filled with helium and
comparing to measurements in vacuum.  
The product $kT$, currently known with an uncertainty of 0.9 ppm \cite{Mohr:12}
near the temperature of the triple point of water, is the subject of
active research and a reduction of this uncertainty can be expected in
near future.
Since $\chi$ is five orders of magnitude smaller than $\alpha$, it can
be computed using the non-relativistic wave function from the expression
$\chi = - e^2\langle r^2\rangle / 3m_{\rm e}c^2  $,   where $e$ and
$m_{\rm e}$ are  the electron charge and mass, $c$ is the speed of
light, and $\langle r^2\rangle$ is the average square of the
electron-nucleus distance.  Also the virial coefficients are known
accurately enough from theory \cite{Cencek:11a,Cencek:12}.  However,
$\alpha$ cannot currently be measured 
with uncertainty lower than 0.2 ppm, so the standard clings upon
theory being able to achieve such accuracy.
This Letter describes calculations of $\alpha$ from
first principles.  We also discuss how $n$ obtained purely from theory
can be used to calibrate refractometers or to correct errors in
interferometric length measurements \cite{Stone:04}. 

Since the radiation frequency of interest,  632.9908 nm 
\cite{Hendricks:14}, is much  smaller than the
lowest resonance,  the frequency dependence of $\alpha$ can be
efficiently calculated from the power series expansion,
$\alpha(\omega) = \alpha_0 + \alpha_2\, \omega^2 + \alpha_4\, \omega^4  
   + \cdots$,
where $\alpha_0$ is the static dipole polarizability.  The coefficients
$\alpha_k$, $k>0$,   describing the frequency dependence of
polarizability, will be referred to as
the (polarizability) dispersion coefficients.
We shall use the atomic units throughout (we never use reduced atomic
units), in particular, $a_0^3$, where $a_0$  is the
bohr radius $\hbar^2/m_e e^2$,  as the unit of  polarizability  and  the
inverse of  the atomic unit of time, $t_0=\hbar^3/m_e e^4$,  as the unit
of  frequency. 
For light systems like helium, each $\alpha_k$ can be expanded in
powers of  the fine structure constant $1/c$,  where
$c$=137.0359991~\cite{Mohr:12} is the speed
of light expressed in atomic units,
$\alpha_k = \alpha_k^{(0)} +  \alpha_k^{(2)} +   \alpha_k^{(3)} + \cdots$,
$\alpha_k^{(l)}$ being  proportional to $1/c^l$.
We shall refer to $\alpha_k^{(2)}$ as the relativistic contributions. 
The terms $\alpha_k^{(3)}$, $\alpha_k^{(4)}$, etc., are due to  radiative 
as well as higher-order relativistic effects predicted by
quantum electrodynamics (QED).

The nuclear mass dependence of  the nonrelativistic polarizability
$\alpha_n^{(0)}$,  can be taken into account exactly, but for the
relativistic and QED  contributions   one has to use an expansion in
powers of   the ratio of $m_{\rm e}$ to the nuclear
mass $m_{\alpha}$, i.e., in powers of
$1/M=m_{\rm e}/m_{\alpha} = 1/7294.2995361$.  Since
$1/M$ is of the order of 10$^{-4}$,  keeping the linear term is entirely
sufficient and such contributions can be represented in the form
$ \alpha_k^{(l)} = \alpha_k^{(l0)} + \alpha_k^{(l1)},\ l \ge 2$,
where $\alpha_k^{(l0)} $ are computed with the infinite nuclear mass and 
$\alpha_k^{(l1)}$ are corrections of the order of $1/ (Mc^l)$, referred to
as the recoil corrections.  These recoil corrections are expected to be
negligible except  for the static ones $ \alpha_0^{(21)}$ and
$\alpha_0^{(31)}$ and, possibly, for $\alpha_2^{(21)}$.

For comparisons with experiments, it is convenient  to convert
frequency to wave length $\lambda = 2\pi c e^2/\hbar\omega$
\begin{equation}\label{def_A_n}
\alpha(\lambda) = A_0 + A_2 \, \lambda^{-2} + A_4 \,\lambda^{-4}    + \cdots .
\end{equation}
When $\alpha(\lambda)$ remains in atomic units and $\lambda$ is measured
in nm, the relation is $A_k = f^k \alpha_k$,
$f = 2\pi c \, a_0/{\rm nm} = 45.56335253$
(with $a_0$=0.05291772109 nm ).    
  
For $\alpha_0$ contributions,
our more accurate values are consistent with Ref.~\onlinecite{Lach:04}
to all digits published 
except for the term describing the electric-field dependence of Bethe's
logarithm and for $\alpha_0^{(40)}$.  We computed the former term
using a new method since Ref.~\onlinecite{Lach:04} was the only source
for this quantity.  The contributions of this component from the two
calculations differ only marginally, by 0.011 $\mu a_0^3$.
The term $\alpha_0^{(40)}$ was estimated in Ref.~\onlinecite{Lach:04}
by the contribution from the simple one-loop
expression \cite{Pachucki:00} and the uncertainty of this term was
assumed to be 40\%.  Later, it was shown in
Ref.~\onlinecite{Pachucki:06b} that the error of one-loop approximation
applied to the excitation energies of helium is only about 5\%.
Therefore, we reduced our estimate from 40\% to 25\% or 0.14 $\mu a_0^3$,
which we believe is still conservative.

The dispersion coefficients  $\alpha_k$ ($k=2,\,4,\,6$) were calculated
thus far only by Bhatia and Drachman (BD) \cite{Bhatia:97,Bhatia:98}.
However, these authors did not provide any estimates of the
uncertainties.  Their relativistic contributions do depend on the nuclear
mass but the recoil effect, $ \alpha_k^{(21)}$,  was not correctly
taken into account, vide infra.
%
Furthermore, the $A_k$ coefficients were incorrectly converted from the
reduced Rydberg units: the factor $(1+m_e/m_\alpha)^k$,
appearing in the correct conversion formula, was erroneously replaced
by  its square $(1+m_e/m_\alpha)^{2k}$. 

 
 
At the nonrelativistic level of theory,
$\alpha(\omega)$  of an atom in a quantum state $\psi$ 
is defined by the standard  polarization propagator expression
\begin{equation}\label{def_alpha}
\alpha(\omega) = 
\langle \psi | z\,{\cal R}(\omega) z| \psi \rangle  
+  \langle \psi | z\,{\RR}(-\omega) z| \psi \rangle   ,
\end{equation}
where $z = z_1 +z_2$, with $z_i$ denoting electron coordinates, and
$ {\cal R}(\omega)  = Q\,(QH -E +\omega)^{-1}$ 
is the resolvent of the atomic Hamiltonian $H$,
with $Q= 1 - P = 1- |\psi\rangle\langle\psi|$ and 
$E$ being the energy of state $\psi$. 
For the helium atom
\begin{equation}\label{H1}
 H  =-\frac{1}{2}{\nabla}_1^2 -\frac{1}{2}{\nabla}_2^2 
     -\frac{1}{2M}({\nabla}_1+{\nabla}_2)^2
     -\frac{2}{r_1}-\frac{2}{r_2}+\frac{1}{r_{12}}.
\end{equation}
$\RR(\omega)$ satisfies the identity  
$\RR(\omega) = \RR -\omega\, \RR \, \RR(\omega)$,
where $\RR =  Q\,(QH -E)^{-1}$ is the static (reduced) resolvent of $H$. 
Iterating this expression
and inserting it into Eq.~(\ref{def_alpha}), one obtains
\begin{equation}\label{an0}
   \alpha^{(0)}_k = 2\, \langle \psi | z\,{\cal R}^{k+1} z| \psi \rangle.
\end{equation} 
 

To account for the leading relativistic contributions of the order of $1/c^2$
assuming infinite nuclear mass, we  
add to the Hamiltonian of Eq. (\ref{H1}) the perturbation
from the Breit-Pauli Hamiltonian \cite{Bethe:57} obtaining
\begin{equation*}\label{a020}
\alpha_0^{(20)} = -4\, \langle \psi_0|B_1\RR_0z\RR_0 z \psi_0\rangle 
                 -2\, \langle \psi_0|z\RR_0\overline{B}_1\RR_0 z \psi_0\rangle ,
\end{equation*}
\begin{eqnarray} \nonumber
\alpha_2^{(20)}& =& -4\,\langle \psi_0|B_1\RR_0z\RR^3_0 z \psi_0\rangle 
                -4\,\langle \psi_0|z\RR_0\overline{B}_1\RR^3_0 z \psi_0\rangle\\
           & &-2\,\langle \psi_0|z\RR^2_0\overline{B}_1\RR^2_0 z \psi_0\rangle ,
                  \nonumber
\end{eqnarray}
 \begin{eqnarray}\label{a420}\nonumber
\alpha_4^{(20)}& =& -4\,\langle \psi_0|B_1\RR_0z\RR^5_0 z \psi_0\rangle 
               -4\,\langle \psi_0|z\RR_0\overline{B}_1\RR^5_0 z \psi_0\rangle\\
                       &    & 
            -4\,\langle \psi_0|z\RR^2 _0\overline{B}_1\RR^4_0 z \psi_0\rangle
            -2\,\langle \psi_0|z\RR^3_0\overline{B}_1\RR^3_0 z \psi_0\rangle ,
                             \nonumber
\end{eqnarray}
 \begin{eqnarray}\label{a620}\nonumber
\alpha_6^{(20)}& =& -4\,\langle \psi_0|B_1\RR_0z\RR^7_0 z \psi_0\rangle 
               -4\,\langle \psi_0|z\RR_0\overline{B}_1\RR^7_0 z \psi_0\rangle\\
                       &    & 
             -4\,\langle \psi_0|z\RR^2 _0\overline{B}_1\RR^6_0 z \psi_0\rangle
    -4\,\langle \psi_0|z\RR^3_0\overline{B}_1\RR^5_0 z \psi_0\rangle 
              \nonumber\\
       & &  -2 \, \langle \psi_0|z\RR^4_0\overline{B}_1\RR^4_0 z \psi_0\rangle ,
              \nonumber
\end{eqnarray}
where
\begin{eqnarray}\label{B1}
B_1 &=& -\frac{1}{8c^2}(\nabla_1^4+\nabla_2^4) +  
\frac{\pi }{ c^2}[\delta (\v{r}_1)+
\delta (\v{r}_2)] +\frac{\pi }{ c^2} \delta (\v{r}_{12})
                                        \nonumber \\
 &&+\frac{1}{2c^2} [{\nabla}_1 r^{-1}_{12} {\nabla}_2
   +({\nabla}_1\v{r}_{12}) r_{12}^{-3} (\v{r}_{12}{\nabla}_2)]  ,
\end{eqnarray}
$\overline{B}_1 =  B_1 - \langle \psi_0 | B_1\psi_0\rangle$, and
the quantities with subscript 0 are analogous to those defined
above but for infinite-mass nonrelativistic Hamiltonian.


The relativistic recoil term
$\alpha_0^{(21)}$ is equal $\alpha_0^{(21)}(B_2) + \alpha_0^{(21)}(H_1B_1)$,
where $H_1 = -\frac{1}{2M}({\nabla}_1+{\nabla}_2)^2$ and
\begin{align}
 B_2 & = \frac{1}{Mc^2}\! \Big[
    {\nabla}_1 r^{-1}_1 {\nabla}_1
  + {({\nabla}_1\v{r}_1){r_1^{-3}}(\v{r}_1{\nabla}_1)}
  +  {{\nabla}_1{r^{-1}_1}{\nabla}_2}
    \nonumber \\
 &+ ({\nabla}_1\v{r}_1){r_1^{-3}(\v{r}_1{\nabla}_2)}
  +   {{\nabla}_2{r^{-1}_2}{\nabla}_1}
  + {({\nabla}_2\v{r}_2){r_2^{-3}}(\v{r}_2{\nabla}_1)}
    \nonumber \\
&  + {{\nabla}_2{r^{-1}_2}{\nabla}_2}
   + {({\nabla}_2\v{r}_2){r_2^{-3}}(\v{r}_2{\nabla}_2)}\Big] .
\end{align}
The two components are given by
\begin{equation}\label{a021L}
  \alpha_0^{(21)}(B_2) =  -4\, \langle \psi_0|B_2\RR_0z\RR_0 z \psi_0\rangle 
               -2\, \langle \psi_0|z\RR_0\overline{B}_2\RR_0 z \psi_0\rangle  
\end{equation}
where $\overline{B}_2$ is analogous to $\overline{B}_1$ and
\begin{align}\label{a021B}
&\alpha_0^{(21)}(H_1 B_1) =
       \nonumber \\
&\quad 4\left[\langle \psi_0| z\RR_0z \RR_0 \overline{H}_1\RR_0B_1\psi_0\rangle
  +\langle \psi_0| z\RR_0z \RR_0 \overline{B}_1\RR_0H_1\psi_0\rangle
   \right. \nonumber \\
&\quad +\langle \psi_0| z\RR_0 \overline{H}_1\RR_0z\RR_0B_1\psi_0\rangle
 +\langle \psi_0| z\RR_0 \overline{B}_1\RR_0z\RR_0H_1\psi_0\rangle
          \nonumber \\
&\quad +\langle\psi_0|z\RR_0\overline{H}_1\RR_0\overline{B}_1\RR_0z\psi_0\rangle
 +\langle \psi_0|H_1\RR_0z\RR_0z\RR_0B_1\psi_0\rangle 
         \nonumber \\
&\quad  -\langle \psi_0| z\RR_0 z\psi_0\rangle 
                \langle \psi_0|H_1\RR^2_0 B_1\psi_0\rangle 
         \nonumber \\
&\quad         \left. -\langle \psi_0| z\RR^2_0 z\psi_0\rangle 
                \langle \psi_0|H_1\RR_0 B_1\psi_0\rangle \right],
\end{align} 
where $\overline{H}_1 =  H_1 - \langle \psi_0 | H_1\psi_0\rangle$.

The correction $\alpha_2^{(21)}$ is very small and can be computed 
using a finite difference expression
\begin{equation}\label{a212}
\alpha_2^{(21)} \approx  \alpha_2^{(20)}(B_1\rightarrow B_2)
                       + \alpha_2^{(20)}(H_0\rightarrow H)
                       - \alpha_2^{(20)} ,
\end{equation}
valid to the order of $1/(M^2c^2)$,
where $B_1\rightarrow B_2$ means that operator $B_1$ in the expression
for $\alpha_2^{(20)}$ should be replaced by $B_2$ and similarly
$H_0\rightarrow H$ means that quantities computed with the Hamiltonian
$H_0$ should be replaced by those computed with $H$.


To evaluate $\alpha_k^{(li)}$,
accurate representations of the helium ground-state
wave functions $\psi_0$ and $\psi$
were obtained by minimizing  the conventional
Rayleigh-Ritz functional for the Hamiltonians $H_0$ and $H$,
respectively. The
auxiliary functions were obtained recursively from Hylleraas-type functionals
\begin{equation}\label{J}
 {J}^{(n)}_0[\tilde{\phi}]=
\langle\tilde{\phi} |H_0-E_0 +P_0 |\tilde{\phi} \rangle    
    -2  \langle\tilde{\phi} |\phi_0^{(n-1)}\rangle
\end{equation}
for $\phi^{(n)}_0 = \RR_0^nz\psi_0$ and
\begin{equation}\label{K}
 {K}_0^{(n)}[\tilde{\psi} ]=
  \langle\tilde{\psi} |H_0-E_0 +P_0 |\tilde{\psi} \rangle
     -2\langle\tilde{\psi} |(1-P_0)  z\phi_0^{(n)}\rangle\, ,
\end{equation}
for $\psi^{(n)}_0 = \RR_0
z\phi^{(n)}_0$, and analogous functionals obtained by
dropping all the subscripts 0 for $\phi^{(n)}  = \RR^n z\psi$ and
$\psi^{(n)}  = \RR z\phi^{(n)}$.
The trial functions used in all minimization processes were expanded
in bases of Slater geminals
\begin{equation}
\tilde{\phi} = (1+{\cal P}_{12})\,  Y({\bf r}_1,{\bf r}_2 ) 
\sum_{i=1}^N c_{i} 
{\rm e}^{-\alpha_i  r_1-\beta_i\, r_2-\gamma_i r_{12}}\, , 
\end{equation}
where ${\cal P}_{12}$ is the transposition operator whereas
$Y({\bf r}_1,{\bf r}_2 ) =z_1$ in calculations of  $\phi^{(n)}_0$  and
$\phi^{(n)} $ and $Y({\bf r}_1,{\bf r}_2 ) =1$ otherwise.
One may note that the functions $\psi^{(n)}_0$
and $\psi^{(n)} $ contain also a $D$ component,  but it
does not contribute to matrix elements that are needed.
The linear coefficients were obtained by solving the appropriate set of
linear equations, while to determine  the nonlinear parameters we
employed two strategies: the full optimization (FO)  and the stochastic
optimization (SO).  In the latter case, the parameters
$\alpha_i$,  $\beta_i$, $\gamma_i$ are pseudo-randomly generated from a box
with optimized dimensions.  We used two boxes to model the short-range
and medium-range asymptotics of the wave functions.  To eliminate
possibilities of numerical errors, the FO and SO based codes (including
the integral and linear algebra routines) were programmed entirely
independently by different members of our team.


The contributions $\alpha^{(0)}_k$ for $k=0,\,2,\,4,\,6$ were computed
for several values of $N$, up to 600 (800) in the FO (SO) approach,
both optimizations giving at least 11 convergent digits, with FO converging
faster.  Our results agree to 9, 8, 4, and 7 digits, respectively, with
the values obtained by  BD \cite{Bhatia:97}.
Using the SO procedure, we also calculated:
$\alpha^{(0)}_{ 8} $=4.39500532(1),  
$\alpha^{(0)}_{10}$=6.7725956(1), 
$\alpha^{(0)}_{12}$=10.622083(1), and 
$\alpha^{(0)}_{14}$=16.86118(1) $\mu a_0^3$.

For the relativistic contributions $\alpha^{(20)}_k$, $k=0,\,2,\,4,\,6$,
the convergence is much slower than in the nonrelativistic case.  This
is due to the fact that we use nonrelativistic functionals which are
sensitive to wave function values in different regions of the
configuration space than the relativistic operators (these operators are
too singular to be used in optimizations).
The SO procedure leads now to a faster convergence than FO since
randomly chosen exponents cover the space more uniformly than FO
exponents. Thus, we used the SO results as our recommended values and in
estimates of uncertainty.  Nevertheless, the agreement to 6, 5, 3, and 3
digits, respectively, between the two sets of results is more than
sufficient for the present purposes.
Our values are substantially more accurate than those of 
BD \cite{Bhatia:98},  with agreement to only
2, 2, and 1 digit, respectively (BD did not compute $\alpha^{(20)}_6$).
For $\alpha^{(20)}_0$, our results are consistent with, but
significantly more accurate than calculations of
Refs.~\cite{Cencek:01,Pachucki:01,Lach:04}.
The relativistic recoil contribution
$\alpha^{(21)}_0$=$-$0.0935(1) $\mu a_0^3$. Its smallness results from
some cancellation of its components,  $\alpha^{(21)}_0(H_1B_1)$ and
$\alpha^{(21)}_0(B_2)$, equal
to 0.1559 and $-$0.2494 $\mu a_0^3$, respectively.
The contribution $\alpha^{(21)}_2$ 
is equal to $-$0.144(1) $\mu a_0^3$, so it is virtually negligible.

It should be pointed out that the relativistic contributions computed by
BD \cite{Bhatia:98} depend on the nuclear mass and,
strictly speaking, should not be compared with our,
nuclear-mass-independent contributions  $\alpha^{(20)}_k$.  This is
because these authors incorrectly assumed that the individual
terms in the Breit-Pauli Hamiltonian are proportional to (inverse)
powers of the reduced electron mass rather than the real mass.
Therefore, although the nuclear-mass-dependent part of their
relativistic contributions is of the order of $1/(Mc^2)$, it differs from
the  $\alpha^{(21)}_k(B_1H_1)$ part of the true recoil correction.
Additionally, BD completely neglected the contribution
$\alpha^{(21)}_k(B_2)$. Thus, their
relativistic contributions cannot be viewed as approximations to
$\alpha^{(20)}_k +\alpha^{(21)}_k$.  Since the effects of the order of
$1/(Mc^2)$ are very small, the  differences between our relativistic
contributions and those of BD are mainly due to the
differences in basis sets used in the calculations rather than to the
treatments of the nuclear mass dependence.    

After correcting the units conversion error in
Ref.~\onlinecite{Bhatia:98} as discussed earlier, the $A_k$ coefficients
computed by BD agree with our values to 5, 6, 4, and 5 digits for $k =
0,\, 2,\, 4,\, 6$, respectively.  For $k=0$, the discrepancy is mainly
due to the $1/c^3$ terms not considered by BD.  The reasons for the
relatively low accuracy of $A_4$ are unclear.   Due to the smallness of
the relativistic contributions to $A_k$, the overall agreement is good
despite the fact that the relativistic contributions from BD work are
significantly less accurate than ours.

\begin{table}[h] 
\caption{Dynamic polarizability of $^4$He [$a_0^3$] at  
         $\lambda$ =  632.9908 nm.}
\label{BD2}

\begin{tabular}{lll} \hline\hline  
static & nonrelativistic          & \qw$1.383\,809\,98641(1)$ \\
       & $1/c^2\mbox{ }^a$        & $-0.000\,080\,4534(1)$ \\
       & $1/c^3\mbox{ }^b$        & \qw$0.000\,030\,655(1)$ \\
       & $1/c^4$                  & \qw$0.000\,000\,56(14)$ \\
       & finite nuclear size$^c$  & \qw$0.000\,000\,0217(1)$ \\
       & total                    & \qw$1.383\,760\,77(14)$  \\ \hline     
$\lambda^{-2}$ & nonrelativistic  & \qw$0.007\,995\,7979(1)$ \\
               & relativistic$^d$ & $-0.000\,000\,1721(1)$ \\
               & total            & \qw$0.007\,995\,6258(1)$ \\ \hline 
$\lambda^{-4}$ & nonrelativistic  & \qw$0.000\,054\,8363(1)$ \\
               & relativistic     & \qw$0.000\,000\,00014(1)$ \\
               & total            & \qw$0.000\,054\,8364(1)$ \\ \hline 
$\lambda^{-6}$ & nonrelativistic  & \qw$0.000\,000\,4076(1)$ \\
               & relativistic     & \qw$0.000\,000\,0000(1)$  \\
               & total            & \qw$0.000\,000\,4076(1)$  \\ \hline 
$\lambda^{-8}$ & nonrelativistic  & \qw$0.000\,000\,0032(1)$  \\ \hline 
$\alpha(\lambda) - \alpha(0)$
               & present$^e$      & \qw$0.008\,050\,8730(1)$  \\
               & BD$^f$           & \qw$0.008\,050\,871 $  \\ \hline 
total          & present          & \qw$1.391\,811\,64(14)$  \\
               & BD$^g$           & \qw$1.391\,780\,800$  \\
\hline \hline
\end{tabular}
\begin{flushleft}
$^a$Includes the recoil correction of the order of $1/(Mc^2)$
equal to $-0.000\,000\,0935(1)$.
~$^b$From Ref.~\onlinecite{Lach:04} except for the contribution
from the electric field derivative of the Bethe logarithm 
equal to $0.000\,000\,182(1)$ \cite{Puchalski:14}.
~$^c$ Computed adding the correction term
$(4/3)\pi\, r^2_{\alpha}\,[\,\delta (\v{r}_1)+ \delta (\v{r}_2)\,]$ to
$H$, where $r_{\alpha}$ = 1.676 fm is the nuclear radius.
~$^d$Including the recoil correction of the order of $1/(Mc^2)$
equal to $-0.000\,000\,00075(1)$.
~$^e$The contribution of the $\lambda^{-10}$ term, amounting to
2.5$\times$10$^{-11}$,  is negligible.
~$^f$Calculated using correctly converted $A_k$ constants.
Equation (15) of Ref.~\onlinecite{Bhatia:98} gives $0.008\,052\,951$,
i.e.,  0.03\% error resulting in 
1.7 ppm error in the total value of $\alpha(632.9908)$.
~$^g$Using the static value of BD equal to $1.383\,729\,929$.
\end{flushleft}
\vspace*{-8mm}
\end{table}

In Table \ref{BD2}, we present the dynamic polarizability of $^4$He.
In addition to the
contributions discussed earlier, we included the effect of finite
nuclear size which is almost negligible.  The dispersion part of
$\alpha(632.9908)$, i.e., the contribution explicitly dependent on
wavelength, agrees to 6 significant digits with the result of BD (after
conversion errors are corrected) due to the high, eight-digit accuracy of
BD's $\alpha_2^{(0)}$ contribution.
However, the total polarizability obtained by us differs significantly,
agreement to 5 digits and discrepancy of about 22 ppm, from BD's result.
As already discussed, this difference is mainly due to the  QED effects
neglected by these authors.  The second  source of the difference is our
significantly improved  value of the static relativistic component.
The uncertainty of our recommended value of $\alpha(632.9908)$ amounts 0.14
$\mu a_0^3$, i.e.,  about 0.1 ppm.  This accuracy is
sufficient for the purpose of the new pressure standard but one should
ask if any neglected effects could contribute above the uncertainty
estimate.  The potential candidates are the QED recoil correction
$\alpha_0^{(31)}$ of the order of $1/(Mc^3)$, the QED contribution  to the
polarizability dispersion $\alpha_2^{(30)}$ of the order of  $1/c^3$
and, finally, and probably most importantly, the remaining, other than
one-loop contributions to $\alpha_0^{(40)}$ of the order of $1/c^4$.  We
believe that such neglected contributions should not contribute more than
0.1 ppm but are investigating such terms.


\begin{table}[h] 
\caption{Virial expansion of refractive index.
$a_r = \frac{2}{3}a_n$, $b_n$, and $c_n$ are in units of 
\mbox{cm}$^3$\!/mol, cm$^6$\!/mol$^2$, and cm$^9$\!/mol$^3$,
respectively, and $\lambda$ is in nm.
\mbox{1 cm}$^3$\!/mol = 11.205\,8721\! $a_0^3$.
}
\label{refrind}
\begin{tabular}{lll} \hline\hline
$a_r(0)$  & present                & $0.517\,246\,21(6)^{a,b}$ \\
                      & exp.~\cite{Schmidt:07} &  0.517\,245\,5(47)\\
$a_r(632.9908)$  \ \ \ \  
                      & present                & $0.520\,255\,64(6)^{c,d}$ \\
                      & exp.~\cite{Achtermann:91,Achtermann:93}
                                               &  0.521\,3(1)\\
                      & exp.~\cite{Birch:91}   &  0.522\,0(3)\\
$a_r(546.2268)$   
                      & present                & $0.521\,297\,25(6)$ \\
                      & exp.~\cite{Leonard:74} &  0.521\,57(15)$^e$ \\
$b_n(0)$   & present, 273.16 K \ \ \  &  0.0245(2)$^{f}$ \\
$b_n(632.9908)$  
                      & present, 273.16 K  &  0.0238(2)$^{f}$ \\
                      & present, 302 K     &  0.0184(2)$^{f}$ \\
                      & present, 323 K     &  0.0151(2)$^{f}$ \\
                      & exp.~\cite{Achtermann:93}
                                               &  0.000(15)$^g$ \\
$c_n(0)$   & present, 273.16 K  &  -0.93(25)$^{f}$ \\
\hline \hline
\end{tabular}
\begin{flushleft}
$^a$ $a_n = $ 8.694\,292\,2(9) $a_0^3$
or  0.775\,869\,31(8)(4)\,cm$^3$/mol, where the second uncertainty
originates from the Avogadro constant
6.022\,141\,29(27)$\times$10$^{23}$.
~$^b$ Computed using $\chi$ = $-$0.000\,021\,194(1) $a_0^3$
\cite{Bruch:02}, the uncertainty reflects the estimated size
of relativistic contributions.
~$^c$ $a_n = $ 8.744\,8758(9)\,$a_0^3$ = 0.780\,383\,35(8)(4) cm$^3$/mol.
~$^d$ We neglected the frequency dependence of $\chi$,  a relativistic
effect expected to be very small.
~$^e$Inferred from measured value of $n$\,--\,1 = 34.895(10)$\times$$10^{-4}$
at $T$=273.16 K and $p$=101.325 kPa.
~$^f$ Computed using $b_{\varepsilon}$=$-$0.0978(2) cm$^3$/mol 
\cite{Rizzo:02} (uncertainty estimated based on comparison
with Ref.~\onlinecite{Cencek:11a}) and $c_{\varepsilon}$ =$-$1.34(36)
cm$^6$/mol$^2$ \cite{Heller:74,Schmidt:07}, $T$=273.16 K.
For other $T$:
$b_{\varepsilon}$(303) = $-$0.1065(2)\,cm$^3$/mol and
$b_{\varepsilon}$(323) = $-$0.1107(2)\,cm$^3$/mol \cite{Rizzo:02}.
~$^g$ Computed from $a_r$ = 0.5213(1) cm$^3$/mol and
$b_r$ = $\frac{2}{3}b_n - \frac{1}{4}a_r$ = $-$0.068(10) cm$^6$/mol$^2$
measured in Ref.~\onlinecite{Achtermann:93}, same value obtained for
both temperatures.
\end{flushleft}
\vspace*{-5mm}
\end{table}

The virial expansion for the refractive index can be written as
\begin{eqnarray}
\label{n} 
n&=&1 + a_n \rho +b_n \rho^2 + c_n \rho^3 +\cdots ,\\
\label{n1}
a_n& =& 2\pi (\alpha+\chi),\\ \label{n2}
b_n& =& 2\pi (\alpha b_{\varepsilon} 
     +{  \tfrac{1}{3} } \pi \alpha^2 +   \chi  b_{\mu} 
     + \tfrac{1}{3}\pi \chi^2 +2\pi \alpha\chi ), \\
                                                  \label{n3}
c_n &=& 2\pi (\alpha c_{\varepsilon}    
     + \tfrac{2}{3}   \pi \alpha^2 \,  b_{\varepsilon}
     + \tfrac{10}{9}\pi^2\alpha^3),
\end{eqnarray}
where $\rho$ is density, $b_{\varepsilon}$ and $c_{\varepsilon}$
are the dielectric virial
coefficients and $b_{\mu}$ is the magnetic permeability virial coefficient.
We have written down the term $\chi b_{\mu}$ in Eq.~(\ref{n2}), but we
will neglect it in numerical calculations
since $\chi$ is about five orders of magnitude smaller than $\alpha$ and
$b_{\mu}$ (unknown) is expected to be at the most of the same
order as $b_{\varepsilon}$.
We completely ignored the magnetic part of $c_n$ in Eq.~(\ref{n3}). 
After Eq.~(\ref{n}) is squared, it becomes consistent with
Eq.~(4) of Ref.~\onlinecite{Schmidt:07} within the terms included there
except that the
factor of 2 is missing in front of the  $A^2_{\varepsilon}\, b \,\rho^2$
term.   Equation (\ref{n}) can be easily
solved for $\rho$ and the resulting formula
can be used for a determination of density, or, when combined with the
virial equation of state, also for a determination of pressure.

The virial coefficients are presented in Table~\ref{refrind}.  The
agreement with the measurement of Schmidt {\em et al.} \cite{Schmidt:07}
is excellent, to within 1.4$\pm$9.1 ppm.  Note that the authors of
Ref.~\onlinecite{Schmidt:07} reported the value of $a_r$ with a subtracted
magnetic contribution  $4\pi\chi /3 =-0.0000080$ cm$^3$/mol, which was
added back in Table~\ref{refrind}.
%
The agreement with
measurements at 632.9908 nm~\cite{Achtermann:91,Achtermann:93,Birch:91}
is, however, poor,
as noticed earlier comparing with
older theoretical results by  BD \cite{Bhatia:98} and
by Stone and Stejskal~\cite{Stone:04}.
The disagreement with the measurement of Leonard~\cite{Leonard:74} is
smaller,  only about twice the experimental uncertainty.  The apparent better
agreement of theory with this experiment (within 1$\sigma$) found in
Ref.~\onlinecite{Bhatia:98} was due to the neglect \cite{Bhatia:98}
of the nonlinear dependence of density on pressure.
The values of $b_n$ and $c_n$ presented in Table~\ref{refrind} have
uncertainties
due entirely to uncertainties of $b_{\varepsilon}$ and
$c_{\varepsilon}$.  The third  and fourth term in Eq.~(\ref{n2}) make
negligible contributions and there is substantial cancellation between
the first two terms.
The experimental $b_n$ determined from the values measured
in Ref.~\cite{Achtermann:93} is consistent with zero, which is almost within
the combined uncertainties.  From our data and from the
$b_\varepsilon(T)$  data of Ref.~\onlinecite{Rizzo:02} we predict
that  $b_n$ will vanish only around 415 K.   
Since $b_n$ is  small at $T$=273.16 K and
higher temperatures, its accuracy is sufficient to predict $n$\,--\,1
with a 1 ppm uncertainty for pressures up to 10 MPa.
   
\begin{acknowledgments}
We thank Drs.~Jay Hendricks and Jack Stone for suggesting the subject
and valuable discussions and Dr. Mike Moldover for comments on
the manuscript.  This work was supported by the NIST
grant 60NANB13D209, NSF grant CHE-1152899, Polish Ministry
of Science grant NN204~182840, and National Science Center grant
2011/01/B/ST4/00733. 
\end{acknowledgments}


\begin{thebibliography}{25}%
\makeatletter
\providecommand \@ifxundefined [1]{%
 \@ifx{#1\undefined}
}%
\providecommand \@ifnum [1]{%
 \ifnum #1\expandafter \@firstoftwo
 \else \expandafter \@secondoftwo
 \fi
}%
\providecommand \@ifx [1]{%
 \ifx #1\expandafter \@firstoftwo
 \else \expandafter \@secondoftwo
 \fi
}%
\providecommand \natexlab [1]{#1}%
\providecommand \enquote  [1]{``#1''}%
\providecommand \bibnamefont  [1]{#1}%
\providecommand \bibfnamefont [1]{#1}%
\providecommand \citenamefont [1]{#1}%
\providecommand \href@noop [0]{\@secondoftwo}%
\providecommand \href [0]{\begingroup \@sanitize@url \@href}%
\providecommand \@href[1]{\@@startlink{#1}\@@href}%
\providecommand \@@href[1]{\endgroup#1\@@endlink}%
\providecommand \@sanitize@url [0]{\catcode `\\12\catcode `\$12\catcode
  `\&12\catcode `\#12\catcode `\^12\catcode `\_12\catcode `\%12\relax}%
\providecommand \@@startlink[1]{}%
\providecommand \@@endlink[0]{}%
\providecommand \url  [0]{\begingroup\@sanitize@url \@url }%
\providecommand \@url [1]{\endgroup\@href {#1}{\urlprefix }}%
\providecommand \urlprefix  [0]{URL }%
\providecommand \Eprint [0]{\href }%
\providecommand \doibase [0]{http://dx.doi.org/}%
\providecommand \selectlanguage [0]{\@gobble}%
\providecommand \bibinfo  [0]{\@secondoftwo}%
\providecommand \bibfield  [0]{\@secondoftwo}%
\providecommand \translation [1]{[#1]}%
\providecommand \BibitemOpen [0]{}%
\providecommand \bibitemStop [0]{}%
\providecommand \bibitemNoStop [0]{.\EOS\space}%
\providecommand \EOS [0]{\spacefactor3000\relax}%
\providecommand \BibitemShut  [1]{\csname bibitem#1\endcsname}%
\let\auto@bib@innerbib\@empty
\bibitem [{\citenamefont {{\L}ach}\ \emph {et~al.}(2004)\citenamefont
  {{\L}ach}, \citenamefont {Jeziorski},\ and\ \citenamefont
  {Szalewicz}}]{Lach:04}%
  \BibitemOpen
  \bibfield  {author} {\bibinfo {author} {\bibfnamefont {G.}~\bibnamefont
  {{\L}ach}}, \bibinfo {author} {\bibfnamefont {B.}~\bibnamefont {Jeziorski}},
  \ and\ \bibinfo {author} {\bibfnamefont {K.}~\bibnamefont {Szalewicz}},\
  }\href@noop {} {\bibfield  {journal} {\bibinfo  {journal} {Phys. Rev. Lett.}\
  }\textbf {\bibinfo {volume} {92}},\ \bibinfo {pages} {233001} (\bibinfo
  {year} {2004})}\BibitemShut {NoStop}%
\bibitem [{\citenamefont {Przybytek}\ \emph {et~al.}(2010)\citenamefont
  {Przybytek}, \citenamefont {Cencek}, \citenamefont {Komasa}, \citenamefont
  {{\L}ach}, \citenamefont {Jeziorski},\ and\ \citenamefont
  {Szalewicz}}]{Przybytek:10}%
  \BibitemOpen
  \bibfield  {author} {\bibinfo {author} {\bibfnamefont {M.}~\bibnamefont
  {Przybytek}}, \bibinfo {author} {\bibfnamefont {W.}~\bibnamefont {Cencek}},
  \bibinfo {author} {\bibfnamefont {J.}~\bibnamefont {Komasa}}, \bibinfo
  {author} {\bibfnamefont {G.}~\bibnamefont {{\L}ach}}, \bibinfo {author}
  {\bibfnamefont {B.}~\bibnamefont {Jeziorski}}, \ and\ \bibinfo {author}
  {\bibfnamefont {K.}~\bibnamefont {Szalewicz}},\ }\href@noop {} {\bibfield
  {journal} {\bibinfo  {journal} {Phys. Rev. Lett.}\ }\textbf {\bibinfo
  {volume} {104}},\ \bibinfo {pages} {183003} (\bibinfo {year}
  {2010})}\BibitemShut {NoStop}%
\bibitem [{\citenamefont {Cencek}\ \emph {et~al.}(2012)\citenamefont {Cencek},
  \citenamefont {Przybytek}, \citenamefont {Komasa}, \citenamefont {Mehl},
  \citenamefont {Jeziorski},\ and\ \citenamefont {Szalewicz}}]{Cencek:12}%
  \BibitemOpen
  \bibfield  {author} {\bibinfo {author} {\bibfnamefont {W.}~\bibnamefont
  {Cencek}}, \bibinfo {author} {\bibfnamefont {M.}~\bibnamefont {Przybytek}},
  \bibinfo {author} {\bibfnamefont {J.}~\bibnamefont {Komasa}}, \bibinfo
  {author} {\bibfnamefont {J.~B.}\ \bibnamefont {Mehl}}, \bibinfo {author}
  {\bibfnamefont {B.}~\bibnamefont {Jeziorski}}, \ and\ \bibinfo {author}
  {\bibfnamefont {K.}~\bibnamefont {Szalewicz}},\ }\href@noop {} {\bibfield
  {journal} {\bibinfo  {journal} {J. Chem. Phys.}\ }\textbf {\bibinfo {volume}
  {136}},\ \bibinfo {pages} {224303} (\bibinfo {year} {2012})}\BibitemShut
  {NoStop}%
\bibitem [{\citenamefont {Moldover}\ \emph {et~al.}(2014)\citenamefont
  {Moldover}, \citenamefont {Gavioso}, \citenamefont {Mehl}, \citenamefont
  {Pitre}, \citenamefont {{de Podesta}},\ and\ \citenamefont
  {Zhang}}]{Moldover:14}%
  \BibitemOpen
  \bibfield  {author} {\bibinfo {author} {\bibfnamefont {M.~R.}\ \bibnamefont
  {Moldover}}, \bibinfo {author} {\bibfnamefont {R.~M.}\ \bibnamefont
  {Gavioso}}, \bibinfo {author} {\bibfnamefont {J.~B.}\ \bibnamefont {Mehl}},
  \bibinfo {author} {\bibfnamefont {L.}~\bibnamefont {Pitre}}, \bibinfo
  {author} {\bibfnamefont {M.}~\bibnamefont {{de Podesta}}}, \ and\ \bibinfo
  {author} {\bibfnamefont {J.~T.}\ \bibnamefont {Zhang}},\ }\href@noop {}
  {\bibfield  {journal} {\bibinfo  {journal} {Metrologia}\ }\textbf {\bibinfo
  {volume} {51}},\ \bibinfo {pages} {R1} (\bibinfo {year} {2014})}\BibitemShut
  {NoStop}%
\bibitem [{\citenamefont {Hendricks}\ \emph {et~al.}(2014)\citenamefont
  {Hendricks}, \citenamefont {Ricker}, \citenamefont {Egan},\ and\
  \citenamefont {Strouse}}]{Hendricks:14a}%
  \BibitemOpen
  \bibfield  {author} {\bibinfo {author} {\bibfnamefont {J.}~\bibnamefont
  {Hendricks}}, \bibinfo {author} {\bibfnamefont {J.}~\bibnamefont {Ricker}},
  \bibinfo {author} {\bibfnamefont {P.}~\bibnamefont {Egan}}, \ and\ \bibinfo
  {author} {\bibfnamefont {G.}~\bibnamefont {Strouse}},\ }\href@noop {}
  {\bibfield  {journal} {\bibinfo  {journal} {Phys. World}\ }\textbf {\bibinfo
  {volume} {67}},\ \bibinfo {pages} {13} (\bibinfo {year} {2014})}\BibitemShut
  {NoStop}%
\bibitem [{\citenamefont {Schmidt}\ \emph {et~al.}(2007)\citenamefont
  {Schmidt}, \citenamefont {Gavioso}, \citenamefont {May},\ and\ \citenamefont
  {Moldover}}]{Schmidt:07}%
  \BibitemOpen
  \bibfield  {author} {\bibinfo {author} {\bibfnamefont {J.~W.}\ \bibnamefont
  {Schmidt}}, \bibinfo {author} {\bibfnamefont {R.~M.}\ \bibnamefont
  {Gavioso}}, \bibinfo {author} {\bibfnamefont {E.~F.}\ \bibnamefont {May}}, \
  and\ \bibinfo {author} {\bibfnamefont {M.~R.}\ \bibnamefont {Moldover}},\
  }\href@noop {} {\bibfield  {journal} {\bibinfo  {journal} {Phys. Rev. Lett.}\
  }\textbf {\bibinfo {volume} {98}},\ \bibinfo {pages} {254504} (\bibinfo
  {year} {2007})}\BibitemShut {NoStop}%
\bibitem [{\citenamefont {Mohr}\ \emph {et~al.}(2012)\citenamefont {Mohr},
  \citenamefont {Taylor},\ and\ \citenamefont {Newell}}]{Mohr:12}%
  \BibitemOpen
  \bibfield  {author} {\bibinfo {author} {\bibfnamefont {P.~J.}\ \bibnamefont
  {Mohr}}, \bibinfo {author} {\bibfnamefont {B.~N.}\ \bibnamefont {Taylor}}, \
  and\ \bibinfo {author} {\bibfnamefont {D.~B.}\ \bibnamefont {Newell}},\
  }\href@noop {} {\bibfield  {journal} {\bibinfo  {journal} {Rev. Mod. Phys.}\
  }\textbf {\bibinfo {volume} {84}},\ \bibinfo {pages} {1527} (\bibinfo {year}
  {2012})}\BibitemShut {NoStop}%
\bibitem [{\citenamefont {Cencek}\ \emph {et~al.}(2011)\citenamefont {Cencek},
  \citenamefont {Komasa},\ and\ \citenamefont {Szalewicz}}]{Cencek:11a}%
  \BibitemOpen
  \bibfield  {author} {\bibinfo {author} {\bibfnamefont {W.}~\bibnamefont
  {Cencek}}, \bibinfo {author} {\bibfnamefont {J.}~\bibnamefont {Komasa}}, \
  and\ \bibinfo {author} {\bibfnamefont {K.}~\bibnamefont {Szalewicz}},\
  }\href@noop {} {\bibfield  {journal} {\bibinfo  {journal} {J. Chem. Phys.}\
  }\textbf {\bibinfo {volume} {135}},\ \bibinfo {pages} {014301} (\bibinfo
  {year} {2011})}\BibitemShut {NoStop}%
\bibitem [{\citenamefont {Stone}\ and\ \citenamefont
  {Stejskal}(2004)}]{Stone:04}%
  \BibitemOpen
  \bibfield  {author} {\bibinfo {author} {\bibfnamefont {J.~A.}\ \bibnamefont
  {Stone}}\ and\ \bibinfo {author} {\bibfnamefont {A.}~\bibnamefont
  {Stejskal}},\ }\href@noop {} {\bibfield  {journal} {\bibinfo  {journal}
  {Metrologia}\ }\textbf {\bibinfo {volume} {41}},\ \bibinfo {pages} {189}
  (\bibinfo {year} {2004})}\BibitemShut {NoStop}%
\bibitem [{\citenamefont {Hendricks}(2014)}]{Hendricks:14}%
  \BibitemOpen
  \bibfield  {author} {\bibinfo {author} {\bibfnamefont {J.~H.}\ \bibnamefont
  {Hendricks}},\ }\href@noop {} {} (\bibinfo {year} {2014}),\ \bibinfo {note}
  {{p}rivate communication}\BibitemShut {NoStop}%
\bibitem [{\citenamefont {Pachucki}(2000)}]{Pachucki:00}%
  \BibitemOpen
  \bibfield  {author} {\bibinfo {author} {\bibfnamefont {K.}~\bibnamefont
  {Pachucki}},\ }\href@noop {} {\bibfield  {journal} {\bibinfo  {journal}
  {Phys. Rev. Lett.}\ }\textbf {\bibinfo {volume} {84}},\ \bibinfo {pages}
  {4651} (\bibinfo {year} {2000})}\BibitemShut {NoStop}%
\bibitem [{\citenamefont {Pachucki}(2006)}]{Pachucki:06b}%
  \BibitemOpen
  \bibfield  {author} {\bibinfo {author} {\bibfnamefont {K.}~\bibnamefont
  {Pachucki}},\ }\href@noop {} {\bibfield  {journal} {\bibinfo  {journal}
  {Phys. Rev. A}\ }\textbf {\bibinfo {volume} {74}},\ \bibinfo {pages} {022512}
  (\bibinfo {year} {2006})}\BibitemShut {NoStop}%
\bibitem [{\citenamefont {Bhatia}\ and\ \citenamefont
  {Drachman}(1997)}]{Bhatia:97}%
  \BibitemOpen
  \bibfield  {author} {\bibinfo {author} {\bibfnamefont {A.~K.}\ \bibnamefont
  {Bhatia}}\ and\ \bibinfo {author} {\bibfnamefont {R.~J.}\ \bibnamefont
  {Drachman}},\ }\href@noop {} {\bibfield  {journal} {\bibinfo  {journal} {Can.
  J. Phys.}\ }\textbf {\bibinfo {volume} {75}},\ \bibinfo {pages} {11}
  (\bibinfo {year} {1997})}\BibitemShut {NoStop}%
\bibitem [{\citenamefont {Bhatia}\ and\ \citenamefont
  {Drachman}(1998)}]{Bhatia:98}%
  \BibitemOpen
  \bibfield  {author} {\bibinfo {author} {\bibfnamefont {A.~K.}\ \bibnamefont
  {Bhatia}}\ and\ \bibinfo {author} {\bibfnamefont {R.~J.}\ \bibnamefont
  {Drachman}},\ }\href@noop {} {\bibfield  {journal} {\bibinfo  {journal}
  {Phys. Rev. A}\ }\textbf {\bibinfo {volume} {58}},\ \bibinfo {pages} {4470}
  (\bibinfo {year} {1998})}\BibitemShut {NoStop}%
\bibitem [{\citenamefont {Bethe}\ and\ \citenamefont
  {Salpeter}(1957)}]{Bethe:57}%
  \BibitemOpen
  \bibfield  {author} {\bibinfo {author} {\bibfnamefont {H.~A.}\ \bibnamefont
  {Bethe}}\ and\ \bibinfo {author} {\bibfnamefont {E.~E.}\ \bibnamefont
  {Salpeter}},\ }\href@noop {} {\emph {\bibinfo {title} {Quantum Mechanics of
  One- and Two-Electron Systems}}}\ (\bibinfo  {publisher} {Springer-Verlag,
  Berlin and New York},\ \bibinfo {year} {1957})\BibitemShut {NoStop}%
\bibitem [{\citenamefont {Cencek}\ \emph {et~al.}(2001)\citenamefont {Cencek},
  \citenamefont {Szalewicz},\ and\ \citenamefont {Jeziorski}}]{Cencek:01}%
  \BibitemOpen
  \bibfield  {author} {\bibinfo {author} {\bibfnamefont {W.}~\bibnamefont
  {Cencek}}, \bibinfo {author} {\bibfnamefont {K.}~\bibnamefont {Szalewicz}}, \
  and\ \bibinfo {author} {\bibfnamefont {B.}~\bibnamefont {Jeziorski}},\
  }\href@noop {} {\bibfield  {journal} {\bibinfo  {journal} {Phys. Rev. Lett.}\
  }\textbf {\bibinfo {volume} {86}},\ \bibinfo {pages} {5675} (\bibinfo {year}
  {2001})}\BibitemShut {NoStop}%
\bibitem [{\citenamefont {Pachucki}\ and\ \citenamefont
  {Sapirstein}(2001)}]{Pachucki:01}%
  \BibitemOpen
  \bibfield  {author} {\bibinfo {author} {\bibfnamefont {K.}~\bibnamefont
  {Pachucki}}\ and\ \bibinfo {author} {\bibfnamefont {J.}~\bibnamefont
  {Sapirstein}},\ }\href@noop {} {\bibfield  {journal} {\bibinfo  {journal}
  {Phys. Rev. A}\ }\textbf {\bibinfo {volume} {63}},\ \bibinfo {pages} {012504}
  (\bibinfo {year} {2001})}\BibitemShut {NoStop}%
\bibitem [{\citenamefont {Puchalski}\ \emph {et~al.}(2014)\citenamefont
  {Puchalski}, \citenamefont {Komasa}, \citenamefont {Jeziorski},\ and\
  \citenamefont {Szalewicz}}]{Puchalski:14}%
  \BibitemOpen
  \bibfield  {author} {\bibinfo {author} {\bibfnamefont {M.}~\bibnamefont
  {Puchalski}}, \bibinfo {author} {\bibfnamefont {J.}~\bibnamefont {Komasa}},
  \bibinfo {author} {\bibfnamefont {B.}~\bibnamefont {Jeziorski}}, \ and\
  \bibinfo {author} {\bibfnamefont {K.}~\bibnamefont {Szalewicz}},\ }\href@noop
  {} {} (\bibinfo {year} {2014}),\ \bibinfo {note} {unpublished
  results}\BibitemShut {NoStop}%
\bibitem [{\citenamefont {Achtermann}\ \emph {et~al.}(1991)\citenamefont
  {Achtermann}, \citenamefont {Magnus},\ and\ \citenamefont
  {Bose}}]{Achtermann:91}%
  \BibitemOpen
  \bibfield  {author} {\bibinfo {author} {\bibfnamefont {H.~J.}\ \bibnamefont
  {Achtermann}}, \bibinfo {author} {\bibfnamefont {G.}~\bibnamefont {Magnus}},
  \ and\ \bibinfo {author} {\bibfnamefont {T.~K.}\ \bibnamefont {Bose}},\
  }\href@noop {} {\bibfield  {journal} {\bibinfo  {journal} {J. Chem. Phys.}\
  }\textbf {\bibinfo {volume} {94}},\ \bibinfo {pages} {5669} (\bibinfo {year}
  {1991})}\BibitemShut {NoStop}%
\bibitem [{\citenamefont {Achtermann}\ \emph {et~al.}(1993)\citenamefont
  {Achtermann}, \citenamefont {Hong}, \citenamefont {Magnus}, \citenamefont
  {Aziz},\ and\ \citenamefont {Slaman}}]{Achtermann:93}%
  \BibitemOpen
  \bibfield  {author} {\bibinfo {author} {\bibfnamefont {H.~J.}\ \bibnamefont
  {Achtermann}}, \bibinfo {author} {\bibfnamefont {J.~G.}\ \bibnamefont
  {Hong}}, \bibinfo {author} {\bibfnamefont {G.}~\bibnamefont {Magnus}},
  \bibinfo {author} {\bibfnamefont {R.~A.}\ \bibnamefont {Aziz}}, \ and\
  \bibinfo {author} {\bibfnamefont {M.~J.}\ \bibnamefont {Slaman}},\
  }\href@noop {} {\bibfield  {journal} {\bibinfo  {journal} {J. Chem. Phys.}\
  }\textbf {\bibinfo {volume} {98}},\ \bibinfo {pages} {2308} (\bibinfo {year}
  {1993})}\BibitemShut {NoStop}%
\bibitem [{\citenamefont {Birch}(1991)}]{Birch:91}%
  \BibitemOpen
  \bibfield  {author} {\bibinfo {author} {\bibfnamefont {K.~P.}\ \bibnamefont
  {Birch}},\ }\href@noop {} {\bibfield  {journal} {\bibinfo  {journal} {J. Opt.
  Soc. Am. A}\ }\textbf {\bibinfo {volume} {8}},\ \bibinfo {pages} {1991}
  (\bibinfo {year} {1991})}\BibitemShut {NoStop}%
\bibitem [{\citenamefont {Leonard}(1974)}]{Leonard:74}%
  \BibitemOpen
  \bibfield  {author} {\bibinfo {author} {\bibfnamefont {P.~J.}\ \bibnamefont
  {Leonard}},\ }\href@noop {} {\bibfield  {journal} {\bibinfo  {journal} {At.
  Data Nucl. Data Tables}\ }\textbf {\bibinfo {volume} {14}},\ \bibinfo {pages}
  {21} (\bibinfo {year} {1974})}\BibitemShut {NoStop}%
\bibitem [{\citenamefont {Bruch}\ and\ \citenamefont
  {Weinhold}(2002)}]{Bruch:02}%
  \BibitemOpen
  \bibfield  {author} {\bibinfo {author} {\bibfnamefont {L.~W.}\ \bibnamefont
  {Bruch}}\ and\ \bibinfo {author} {\bibfnamefont {F.}~\bibnamefont
  {Weinhold}},\ }\href@noop {} {\bibfield  {journal} {\bibinfo  {journal} {J.
  Chem. Phys.}\ }\textbf {\bibinfo {volume} {117}},\ \bibinfo {pages} {3243}
  (\bibinfo {year} {2002})},\ \bibinfo {note} {errata: {\bf 119}, 638
  (2003)}\BibitemShut {NoStop}%
\bibitem [{\citenamefont {Rizzo}\ \emph {et~al.}(2002)\citenamefont {Rizzo},
  \citenamefont {H{\"a}ttig}, \citenamefont {Fernandez},\ and\ \citenamefont
  {Koch}}]{Rizzo:02}%
  \BibitemOpen
  \bibfield  {author} {\bibinfo {author} {\bibfnamefont {A.}~\bibnamefont
  {Rizzo}}, \bibinfo {author} {\bibfnamefont {C.}~\bibnamefont {H{\"a}ttig}},
  \bibinfo {author} {\bibfnamefont {B.}~\bibnamefont {Fernandez}}, \ and\
  \bibinfo {author} {\bibfnamefont {H.}~\bibnamefont {Koch}},\ }\href@noop {}
  {\bibfield  {journal} {\bibinfo  {journal} {J. Chem. Phys.}\ }\textbf
  {\bibinfo {volume} {117}},\ \bibinfo {pages} {2609} (\bibinfo {year}
  {2002})}\BibitemShut {NoStop}%
\bibitem [{\citenamefont {Heller}\ and\ \citenamefont
  {Gelbart}(1974)}]{Heller:74}%
  \BibitemOpen
  \bibfield  {author} {\bibinfo {author} {\bibfnamefont {D.~F.}\ \bibnamefont
  {Heller}}\ and\ \bibinfo {author} {\bibfnamefont {W.~M.}\ \bibnamefont
  {Gelbart}},\ }\href@noop {} {\bibfield  {journal} {\bibinfo  {journal} {Chem.
  Phys. Lett.}\ }\textbf {\bibinfo {volume} {27}},\ \bibinfo {pages} {359}
  (\bibinfo {year} {1974})}\BibitemShut {NoStop}%
\end{thebibliography}
\end{document}